 \documentclass[12pt,showpacs,preprintnumbers,floatfix]{article}
\usepackage{graphicx} 
\usepackage{dcolumn} 
\usepackage{bm} 
\usepackage{amsfonts}
\usepackage{amsthm}
\usepackage{amsmath}
\usepackage{amssymb}
\usepackage{epsfig}
\usepackage{color}
\usepackage{textcomp}

\def\be{\begin{equation}}
\def\ee{\end{equation}}
\def\ba{\begin{eqnarray}}
\def\ea{\end{eqnarray}}

\def\bdm{\begin{displaymath}}
\def\edm{\end{displaymath}}
\def\la{~\mbox{\raisebox{-.6ex}{$\stackrel{<}{\sim}$}}~}
\def\ga{~\mbox{\raisebox{-.6ex}{$\stackrel{>}{\sim}$}}~}
\def\bq{\begin{quote}}
\def\eq{\end{quote}}

 at 10truept

\newcommand{\bea}{\begin{eqnarray}}
\newcommand{\eea}{\end{eqnarray}}

\newcommand{\bi}{\begin{itemize}}
\newcommand{\ei}{\end{itemize}}

\newcommand{\beq}{\begin{equation}}
\newcommand{\eeq}{\end{equation}}
\newcommand{\beqa}{\begin{eqnarray}}
\newcommand{\eeqa}{\end{eqnarray}}

\def\la{~\mbox{\raisebox{-.6ex}{$\stackrel{<}{\sim}$}}~}
\def\ga{~\mbox{\raisebox{-.6ex}{$\stackrel{>}{\sim}$}}~}

\def\ltap{\ \raise.3ex\hbox{$<$\kern-.75em\lower1ex\hbox{$\sim$}}\ }
\def\gtap{\ \raise.3ex\hbox{$>$\kern-.75em\lower1ex\hbox{$\sim$}}\ }
\def\gl{\ \raise.5ex\hbox{$>$}\kern-.8em\lower.5ex\hbox{$<$}\ }
\def\roughly#1{\raise.3ex\hbox{$#1$\kern-.75em\lower1ex\hbox{$\sim$}}}

\begin{document}

\thispagestyle{empty}
\begin{flushright}
October 2013
\end{flushright}
\vspace*{1.7cm}
\begin{center}
{\Large \bf Spherical Cows in the Sky with Fab Four}\\

\vspace*{1.2cm} {\large Nemanja Kaloper$^{a,}$\footnote{\tt
kaloper@physics.ucdavis.edu} and McCullen Sandora$^{a,}$\footnote{\tt
mesandora@ucdavis.edu}}\\
\vspace{.5cm} {\em $^a$Department of Physics, University of
California, Davis, CA 95616}\\

\vspace{2cm} ABSTRACT
\end{center}
We explore spherically symmetric static solutions in a subclass of unitary scalar-tensor theories of gravity, called the `Fab Four'
models. The weak field large distance solutions may be phenomenologically viable, but only if the Gauss-Bonnet term is negligible. Only in this limit will the Vainshtein mechanism work consistently. 
Further, classical constraints and unitarity bounds constrain the models quite tightly. Nevertheless, in the limits where the range of individual terms at large scales is respectively Kinetic Braiding,  Horndeski, and 
Gauss-Bonnet, the horizon scale effects may occur while the theory satisfies Solar system constraints and, marginally, unitarity bounds. On the other hand, to bring the cutoff down to below a millimeter constrains all the couplings scales such that `Fab Fours'  can't be heard outside of the Solar system.

\vfill \setcounter{page}{0} \setcounter{footnote}{0}
\newpage

\section{Introduction}

Observation of accelerated expansion of the universe has caused much speculation that
General Relativity (GR) might be modified in the infrared. In general any such modifications
imply the presence of new light degrees of freedom \cite{hkt},\cite{weini}. The simplest 
modifications involve extra scalars, and while some are trivial rewritings of generalized Brans-Dicke 
theories in terms of more obscure variables\footnote{An example of trivial rewritings are the so called ``$f(R)$ gravities".}, 
there are more interesting models which still
obey the usual Cauchy problem formulation of dynamics, based on second order field equations,
which are free of perturbative ghosts and other pathologies (for a recent review, see \cite{tonyco}).
A very general construction of such unitary generalized scalar-tensor theories has been given
by Horndeski in the 70's \cite{horn} (see also \cite{dgsz}). A subclass of Horndeski's theories has been 
identified recently \cite{fab} by cosmological considerations, defined by a Lagrangian
consisting of 4 operators: the usual Einstein-Hilbert term, the Gauss-Bonnet term, the 
so-called `braiding' term, and the double Hodge-dualized Riemann term, all appropriately contracted 
with scalar field derivatives. 

This Lagrangian's complicated structure opens the door for a rich phenomenology
which is still largely unexplored. In particular, the compatibility of this theory with 
solar system tests of gravity has not yet been fully tested. The presence of extra light modes
and extra couplings could yield appreciable corrections to long range gravitational interactions, which are
very strongly constrained by the precision solar system
tests. On the other hand, in such setups environmental suppression of extra forces due to the nonlinear mixing 
generated by extra couplings known as  Vainshtein shielding \cite{vain} may be operational. 
Such phenomena are familiar in other models with higher derivative operators
in the Lagrangians, particularly in massive gravity \cite{vain,babi,sbisniz}, and in 
in galileon theories \cite{gal,hair}.  

In this note we will explore the Vainshtein shielding in the `Fab Four' models, 
which needs to reproduce the standard Newton's law of gravity in the experimentally tested regime of scales.
Clearly, the presence of the many terms in the Lagrangian yields a variety of possible effects 
which should to be constrained. If we neglect the irrelevant operators of the `Fab Four' setup, the strongest
bounds on the couplings of the extra scalar come from the standard Brans-Dicke type scalar force\footnote{Without a hard scalar potential the chameleon effects are irrelevant \cite{cameo}.} \cite{wagoner,damour}, and require adding terms $\propto (\partial \phi)^2$ in order to satisify  
them. However when the extra `Fab Four' terms are turned on, these bounds can be relaxed by the Vainshtein effect, since in much of the phase space the irrelevant operators could be important.

Our strategy is as follows. Using the Brans-Dicke coupling far away to set up the long range fields, we will 
look for the regimes where the Vainshtein shielding kicks in at shorter distances, suppressing the Brans-Dicke corrections to the Newton's law at the scales of the Solar system. We will treat the scalar hair as a perturbation on top of the standard GR background, checking for the limits on parameters that ensure it.  We will also focus on a (very general!) class of models where the coupling functions are analytic in $\phi$, and, consistent with our approximations, retain only the leading order terms in the expansion of the potential functions. This simplifies our analysis since
we can work on a Schwarzschild background, treating both the scalar and the leading nonlinearities as a perturbation. We will ignore the special classes of models where the coupling scales are either $\phi$-independent, have expansion which start with higher powers of $\phi$, and/or involve backgrounds with very large scalar gradients. One expects that these cases either have irrelevantly small effects due to the mixings, or will generically produce bounds from a larger backreaction on the leading order gravitational potential, but a more detailed analysis is beyond the scope of this work.

We will show that, with out restrictions on `Fab Fours', having a successful classical phenomenology requires that the Gauss-Bonnet term is completely marginalized, having no influence on the
dynamics in the linear regime. Further we will consider the bounds that follow from requiring the theory to be perturbative down to experimentally tested scales. We will show that typical `Fab Four' models are either not very interesting cosmologically, because of either classical constraints or unitarity bounds. Even so, if there is a hierarchy between the couplings, allowing the extra operators in `Fab Four' setups (from hereon `extras') to have long range effects at Hubble scale implies that the
perturbation theory breaks down at distances $\sim 1-10$ cm, like in galileon models \cite{gal}. Strictly speaking, if we want to maintain perturbativity down to 
$\sim 1$ mm,  the `Fab Four' extras would be essentially irrelevant beyond the Solar system. 
Nevertheless, such setups might be useful `straw men' for improving the limits on the deviations from General Relativity.

\section{The Framework}

The `Fab Four' action is
\be
S=\int d^4x
\sqrt{\hat g} \Bigl\{\hat V(\varphi) \hat R+\hat V_{GB}(\varphi) \hat G+\hat V_{KB}(\varphi) \hat G^{\mu\nu}\varphi_\mu\varphi_\nu
+\hat V_{H}(\varphi) \hat P^{\alpha\beta}{}_{\mu\nu}\varphi^{\mu}{}_{\alpha}\varphi^\nu\varphi_\beta - {\cal L}_{matter}(\hat g^{\mu\nu}) \Bigr\} \, .
\label{action}
\ee
Here $\varphi$ is a scalar, the indices denote covariant derivatives, $\hat R$ is the Ricci
scalar, $\hat G = \hat R_{\mu\nu\sigma\lambda} \hat R^{\mu\nu\sigma\lambda} - 4
\hat R_{\mu\nu} \hat R^{\mu\nu} + \hat R^2$ 
the Gauss-Bonnet scalar, and $\hat P^{\alpha\beta}{}_{\mu\nu}=\epsilon^{\alpha\beta\gamma\delta} \hat R_{\gamma\delta}{}^{\sigma\lambda}\epsilon_{\sigma\lambda\mu\nu} $ the double dual of the Riemann tensor.  These terms will be referred
to as the Brans-Dicke term, the Gauss-Bonnet term, the kinetic braiding term, and the Horndeski term.  
The latter three terms are covariant extensions to Galileon theories, as found in \cite{dde,aco}.  

We shall look for 
approximate spherically symmetric solutions of the theory (\ref{action}). To this end, we
will use a field redefinition $V(\varphi)\rightarrow\varphi$ to put the first term in
canonical Jordan form, which is allowed in the regime where $V(\varphi)$ is
monotonic\footnote{If $V$ were a constant, $\phi$ and matter would have been trivially decoupled. So we assume that $\partial_\varphi V \ne 0$.}.   
We will then perform a conformal transformation to the Einstein frame, and use the direct scalar-matter coupling as the source of the scalar long range force. This yields the Brans-Dicke scalar hair, which far away is comparable to the Newton's potential of the source. Then we will look for the conditions  that the nonlinear metric-scalar mixings induced by the `extras' give rise to the Vainshtein effect, which splinters the scalar hair close to the source and suppresses the scalar force. At the technical level, this means that we will approximate the 
background in this regime by a linearized Schwarzschild metric, with a subleading static spherically symmetric scalar field on top of it. This approximation requires that energy density in the scalar field (if the Vainshtein mechanism is to occur!)
should be smaller than the ``energy density'' stored in the metric response to the matter source
all the way down to the Schwarzschild radius of the spherical body, where nonlinear effects spoil the Newtonian 
approximation anyway. Checking that this does occur provides extra constraints on the theory.

Performing these transformations, we find the effective action
\ba
S&=& M_{Pl}^2 \int d^4x\sqrt{g} \Bigl\{R-6\partial\phi^2  
-  e^{-4 \phi} {\cal L}_{matter}( e^{2\phi} g^{\mu\nu}) ~~~~~~~~~~~~~~~  \nonumber\\
&& \, \, ~~~~ +V_{GB}(\phi)\Bigl[G+8
G_{\mu\nu}\phi^{\mu\nu}-8R_{\mu\nu}\phi^\mu\phi^\nu-8\phi^{\mu\nu}\phi_{\mu\nu}
+8\Box\phi^2+16\phi_{\mu\nu}\phi^\mu\phi^\nu+8\Box\phi\partial\phi^2\Bigr]\nonumber\\
&& ~~~~~ +V_{KB}(\phi)\Bigl[G_{\mu\nu}-2\phi_{\mu\nu}+2g_{\mu\nu}\Box\phi+3\phi_\mu\phi_\nu\Bigr]\phi^\mu\phi^\nu\nonumber\\
&& \, \, ~~~~ +V_{H}(\phi) \, ~ \Bigl[P^{\alpha\beta}{}_{\mu\nu}
\phi^\mu{}_\alpha\phi^\nu\phi_\beta+\epsilon^2(\phi^\mu\phi_\nu,\phi^\mu{}_\nu)-\partial\phi^2\epsilon(\phi^\mu\phi_\nu,\phi^\mu{}_\nu) \Bigr]
\Bigr\} \, ,
\label{canact}
\ea
where we have field-redefined the scalar according to $\varphi = M_{Pl}^2 \exp({2\phi})$, changed the metric to
$\bar g_{\mu\nu} = \frac{\varphi}{M_{Pl}^2} g_{\mu\nu}$, and reabsorbed the conformal factors into $\hat V_k(\varphi)\rightarrow M_{Pl}^2 V_{k}(\phi)$.  Here we are using the compact notation 
$\epsilon^2(A^\mu_{1\nu},...,A^\mu_{n\nu})=\epsilon_{\mu_1...\mu_n\alpha_{(n+1)}
...\alpha_4}\epsilon_{\nu_1...\nu_n}^{\phantom{\nu_1...\nu_n}\alpha_{(n+1)}
...\alpha_4}A_1^{\mu_1\nu_1}...A_n^{\mu_n\nu_n}$. 
We could now write down the field equations by varying (\ref{canact}) \cite{fab} (which extend the variational equations found in \cite{cdko}). 
Instead we will work in the action, following the logic of \cite{gandalf}. 
So, with our approximations in mind, taking the Schwarzschild background\footnote{We will discuss the validity of this approximation in more detail in
Sec. 3. Among other limitations, this prescription also fails if the energy density in $\phi$ field can distort the Schwarszchild geometry.},
\be
ds^2=-(1-\frac{r_{Schwarzschild}}{r})dt^2+\frac{dr^2}{1-\frac{r_{Schwarzschild}}{r}}+r^2d\Omega_2 \, , \label{rad} 
\ee
we require that $\phi$ only has radial dependence, so that
\be
\phi^\mu{}_\nu=\texttt{diag}\bigg(0,\phi'',\frac{1}{r}\phi',\frac{1}{r}\phi'\bigg)\bigg(1+\mathcal{O}\Big(\frac{r_{Schwarzschild}}{r}\Big)\bigg) \, .
\label{phiderivs}
\ee
Next, we substitute (\ref{rad})-(\ref{phiderivs}) into the action (\ref{canact}), 
and focus on the effects induced by the `Fab Four' extras, specifically looking when the Vainshtein effect works consistent with the linear truncation. Since the Ricci tensor vanishes on the Schwarzschild background, the action 
(\ref{canact}) simplifies dramatically. We find that the leading order scalar Lagrangian in this limit is
\ba
S&=& M_{Pl}^2 \int
drr^2\bigg\{-6\phi'^2+ \frac{ \zeta}{M^2_{Pl}} \phi T  + V_{GB}(\phi)\Big(\frac{r_s^2}{r^6}+\frac{4}{r}
\phi'^3+2\phi'^4\Big)+V_{KB}(\phi)\Big(\frac{4}{r}\phi'^3+3\phi'^4\Big) \nonumber \\
&& \, ~~~~~~~~~~~~~~~~~ + V_{H}(\phi)\Big(\frac{r_s}{r^4}\phi'^3+\frac{2}{r^2}\phi'^4-\frac{2}{r}
\phi'^5\Big)\bigg\} \, .
\label{scalaract}
\ea
The first two terms are the standard Brans-Dicke pieces, where $T$ is the trace of matter stress-energy tensor 
and $\zeta$ is a dimensionless number which is ${\cal O}(1)$ in the `Fab Four' models.
Its precise numerical value is irrelevant, and in what follows we will ignore it; all it does is renormalize the extra scalar force by an ${\cal O}(1)$ factor. As we mentioned, without the Vainshtein effect 
the theory would run afoul of the bounds on deviations from GR. Indeed, as discussed in \cite{hair}, the scalar contribution to the 
force between the source and a probe of unit mass is $\Delta F \simeq \phi'$, with ${\cal O}(1)$ prefactor due to $\zeta/12 \simeq {\cal O}(1)$.
So the Vainshtein effect is necessary to hide this extra force in the regimes where bounds have been obtained, namely, within the Solar system.

We note a few differences between this action and a standard galileon theory. First, 
there is the $\phi'$-independent term in the Gauss-Bonnet
piece, which acts as an additional source, that's absent in the standard galileon setups.  
Next, both the Gauss-Bonnet and
kinetic braiding terms include $\phi'^4$ terms with no $1/r$ prefactors, a
consequence of the presence of $\partial\phi^4$ terms. Lastly, there is a term in the Horndeski part of the action
proportional to $r_s$. It becomes relevant at scales below the Vainshtein
radius, as we shall see below.  

Next, we take the coupling functions $V_k(\phi)$ to be analytic, with 
a Taylor expansion\footnote{This follows since $\phi$ is dimensionless. The factor of $12$ compensates the noncanonical normalization of the $\phi$ kinetic term.} $V_k(\phi) = 12 L_k^{d_k} \sum c_n \phi^n$   where $L_k$ are length scales, $d_k$ the engineering dimension of $V_k$ and
all $c_n \sim {\cal O}(1)$. Hence the leading terms will be the only ones relevant to the leading order, and we can take 
the coupling functions to be constants, except for the Gauss-Bonnet term, where the leading term in the expansion of $V_{GB}$ is a total derivative. 
So to keep the Gauss-Bonnet term operational, we retain the linear term in the expansion. As we will see below, this sources a host of problems with phenomenology, and one of our main conclusions will be that this term must be essentially zero, so that the Gauss-Bonnet coupling function
must start from the quadratic term. We should note here that our logic of truncating the coupling functions to their lowest order terms covers a very broad subclass of the 
`Fab Four' models, and follows naturally from analyticity in $\phi$, which is bounded by its Newtonian value $\phi<r_s/r<1$ at large distances. This does ignore the more restrictivel classes of models where the coupling scales are either $\phi$-independent, have expansion which start with higher powers of $\phi$, and/or involve very large scalar gradients. In these cases the mixing would be either negligible, or the backreaction on the leading order potential would be stronger. A more detailed analysis of these situations is beyond the scope of our paper. One could consider even more general Horndeski theories, involving the fifth `extra', i.e. $\omega (\partial \varphi)^2$ in (\ref{action}), and also 
the galileon terms. In the presence of such terms, the infrared behavior can be significantly modified, further suppressing the forces at Solar system scales, possibly relaxing some of our bounds. 
 
With these assumption we can integrate the scalar field equation once, thanks to the emerging galilean symmetries of the `Fab Four' extras
in this limit.  This enhanced symmetry keeps the neglected terms irrelevant down to the Schwarzschild
radius of the source, if the Vainshtein effect is operational (which demands that $\phi<r_{Schwarzschild}/r$), and so all of the relevant 
physics can be encapsulated by this truncation. Besides making the scalar field equation exactly integrable, galilean symmetry
also allows us to linearly superpose spherical solutions, even if the two sources are
within the nonlinear regime of each other \cite{equiprimp} (so long as the energy density of the superposition is not great enough to source additional nonlinearities \cite{2body}).  This is a
tremendous benefit, since it ensures that the Vainshtein effect works 
for any configuration of compact sources provided that we can show it is
present for a single one in isolation.  The scalar field equation becomes
\be
\phi'+L_{GB}^2\bigg(\frac{12}{r}\phi'^2+8\phi'^3-\frac{r_s^2}{r^5}\bigg)+L_{KB}
^2\bigg(\frac{12}{r}\phi'^2+12\phi'^3\bigg)+L_{H}^4\bigg(\frac{3r_s}{r^4}
\phi'^2+\frac{8}{r^2}\phi'^3-\frac{10}{r}\phi'^4\bigg)=\frac{r_s}{r^2} \, .
\label{scalareom}
\ee 
The source term follows from the Brans-Dicke coupling $\sim \phi T/M^2_{Pl}$, since for static spherically symmetric sources 
$T/M^2_{Pl} =  4 \pi r_{Schwarzschild} \,  \delta^3(\vec r)$ and 
$r_s = {\cal O}(1) \times r_{Schwarzschild}$, accounting for the fact that the parameter $\zeta/12$ is absorbed into it. Because of this in what follows we will not make further distinction between these two length scales. The signs of the extra terms could be either  positive or negative, and the powers are there only to keep track of units. We will focus only on the signs which allow the existence of static spherically symmetric solutions and the Vainshtein shields.  Before presenting
these in full generality, we first consider their individual effects, to better understand the underlying physics.

\section{Vainshtein: Mostly Shielding}

\indent  {\it The Gauss-Bonnet Term}: If the Gauss-Bonnet piece is the only one present we have 
\be
\phi'+\frac{12L_{GB}^2}{r}\phi'^2+8L_{GB}^2\phi'^3
=\frac{r_s}{r^2}\bigg(1+\frac{r_sL_{GB}^2}{
r^3}\bigg) \, .
\label{gbterms}
\ee
Note the
presence of an effective source term on the right hand side.  This equation can be solved for $\phi'$ exactly, 
but in asymptotic regions the
physics is easily understandable.  Above $r_3=(L_{GB}^2r_s)^{1/3}$ the higher order terms are irrelevant, so
we recover $\phi'\rightarrow r_s/r^2$.  Below the scale $r_3$, however, both the
quadratic term in $\phi'$ and the Gauss-Bonnet source are dominant influences,
leading to the solution $\phi'\rightarrow r_s/r^2$ - as it was far away! Hence this completely debilitates 
the Vainshtein mechanism, restoring a Newtonian-like 
profile for $\phi'$ at both small and large length scales, and correcting the standard Newton's force by ${\cal O}(1)$ effects! This
completely invalidates the claims of \cite{aco}: we conclude that the
Gauss-Bonnet piece cannot be allowed to dominate in the regimes where one wants to pass the phenomenological limits.

\indent  {\it The Kinetic Braiding Term}:  Here we set $L_{GB}=L_{H}=0$ and keep only the kinetic braiding effects. The
field equation (\ref{scalareom}) becomes
\be
\phi'+\frac{12L_{KB}^2}{r}\phi'^2+12L^2\phi'^3=\frac{r_s}{r^2} \, .
\label{scalkbeom}
\ee
Again, this can be solved exactly for $\phi'$, but to gain physical insight let's again consider
the asimptotia. We assume (and check {\it a posteriori}) that
the solution is not self-accelerating, so in any given regime one of the terms on
the left hand is dominant, and approximately equal to the source. Far from
the source the $\phi'$ dominates, so a regular Newtonian-like
profile is recovered.  Below the scale $r_3\equiv(r_sL_{KB}^2)^{1/3}$ the quadratic
term dominates, so the scalar profile is 
$\phi'\rightarrow\frac{1}{L_{KB}}\sqrt{\frac{r_s}{r}}$.  The cubic
term is always subdominant, in contrast with the galileon theory, in which the
cubic term dominates the Vainshtein behavior.  This is a consequence of the
quartic single derivative term in the action, which excludes the extra factors
of $L_{KB}/r$ that would normally cause the higher order terms to be more relevant. 
If the cubic terms is taken as a truly small perturbation, the analysis of this
theory was already performed in \cite{hair}. The bottom line is that $L_{KB}$ in this case must be large enough so that 
$r_3$ is bigger than the Solar system size. Taking $r_3 \ga 100$ AU then implies $L_{KB} \ga 10^7$ AU.

\indent  {\it The Horndeski Term}: in the limit where the Horndeski term dominates we find
\be
\phi'+\frac{3L_H^4r_s}{r^4}\phi'^2+\frac{8L_H^4}{r^2}\phi'^3-\frac{10L_H^4}{r}
\phi'^4=\frac{r_s}{r^2} \, .
\label{hornd}
\ee
Here again the far field profile looks asymptotically Newtonian, but below the
scale $r_3$ nonlinearities kick in.  Due to the additional factor of $(r_3/r)^3$
in the quadratic term, the profile tends to $\phi'\rightarrow r/L_H^2$ close to
the source.  This profile is very interesting: not only are all effects
more suppressed than in the standard galileon theories, but also this special radial dependence does not contribute any additional
precession to orbiting bodies! Indeed, the point is that while the general corrections to Newton's potential yield precession of perihelion of 
orbiting bodies\cite{moon,lue,hair}, by Bertrand's theorem the linear potential and the inverse square law do not.  
Thus, only including this term results in what
we might dub \textit{ultra-Vainshtein effect}.  A curiosity is that this profile 
does not depend on the source mass close to the source. But this yields an interesting {\it lower} bound on the length scale $L_H$: in the Solar system
this term would correct the long range potential by a harmonic oscillator-like term. This can be interpreted as being due to additional 
uniformly distributed mass inside the radius at which the potential is being measured. In the Solar system, this mass is bounded by Kepler's laws: it cannot exceed about $10^{-6}$ of the Sun's mass in the region the size of the orbit of Uranus \cite{dicusetal}. So the extra effective long range potential \cite{hair} due to the scalar is 
limited by $\Delta_\phi V \simeq \frac{r^2}{L_H^2} \la 10^{-6} \frac{M_\odot}{M_{Pl}^2 r}$. A quick calculation shows that this implies
$L_H^2 \ga 10^6 \frac{M_{Pl}^2 R_U^3}{M_\odot}$. Using $R_U \simeq 10$ AU and $M_\odot \simeq 10^{57}$ GeV, this yields  $L_H \ga 10^7 {\rm AU}$ which is actually comparable to the bound coming from demanding that the transition to the Vainshtein regime is outside the Solar system. 

\indent  {\it The `Fab Four' in Concert}: 
Combined effects of multiple terms complicate the analysis.  However, the Gauss-Bonnet term always dominates the source below the scale
$r_{3GB}$, yielding the scalar profile $\sim r^{-1}$. To suppress this effect, either $L_{GB}$ must be small, or other terms must dominate in the intervening regions before $r_{3GB}$. Between the Horndeski term and the Kinetic Braiding term, the former is the one that controls the scalar profile closer to the source\footnote{As we will see, the Kinetic Braiding is more important in setting the environmental suppression of the unitarity bound saturation, but we postpone this discussion until the next section.}. The
Kinetic Braiding term cannot win over Gauss-Bonnet term by itself, due to the Gauss-Bonnet sourcing the background, and both of these terms being $\propto \phi'^2/r$: when $r_{3KB}>r_{3GB}$,  for $r$s in this interval 
the Vainshtein shielding suppresses $\phi'$, but stops for $r < r_{3GB}$.  If $r_{3KB}<r_{3GB}$, the Vainshtein shield never raises to begin with. So in the classical limit the Gauss-Bonnet term long range effects can only be suppressed by the Horndeski term, unless $L_{GB}$ is so small that 
its effects do not show beyond the sub-millimeter scales\footnote{A quick and dirty estimate could be found by suppressing $r_{3GB} = (r_s L_{GB}^2)^{1/3}$ to be smaller than about $0.1$ mm for table-top experiments. This is not quite correct, actually, since even if the Gauss-Bonnet begins to influence the fields at larger distances, if the regime dominated by the Kinetic Braiding lasted long enough, the extra $1/r$ 
force could be small. However, in what follows we will obtain a much stronger constraint on the Gauss-Bonnet from considering its backreaction on the background. }. 

Let us  now quantify this. Let's first ignore the Kinetic Braiding term. In this case 
the field equation is
\be
\phi'+\bigg(\frac{12L_{GB}^2}{r}+\frac{3r_sL_H^4}{r^4}\bigg)\phi'^2 + \ldots =\frac{r_s}{
r^2}+\frac{r_s^2L_{GB}^2}{r^5} \, ,
\label{twofabs}
\ee
with the ellipsis denoting the subdominant terms. There are
two crossover scales: $r_{3H}$ and $r_{3GB}$.  Above both, the scalar has a Newtonian profile, as dictated by the source. 
Below both of them we always have
$\phi'\rightarrow \frac{L_{GB}}{L_H^2}\sqrt{\frac{r_s}{r}}$, and the region between the
two depends on whether $L_{GB}$ is greater than or less than $L_H$.  If smaller, the
Horndeski term begins to dominate early, and so there is a region of ultra-Vainshtein effect
before returning to the tamer $r^{-1/2}$ behavior.  
If $L_{GB}>L_H$
the Gauss-Bonnet term will kick in before the Horndeski term, but this just
means that the field profile is unaffected.  The general behavior is that the
only real crossover scale is the one associated with the Horndeski term, and
there is an extra factor of $L_{GB}/L_H$ multiplying the near field profile.  

Now let us suppose that $L_{KB} \gg L_H$ (and ignore the Gauss-Bonnet term which is clearly problematic; this really means we take the limit $L_{KB} \gg L_H \gg L_{GB}$.). In this case, we again find several regimes. Far away, the scalar field has its Brans-Dicke profile $\varphi'  = r_s/r^2$. This persists down to 
$r \sim r_{3KB}$. At distances below $r_{3KB}$, the scalar profile changes to the profile set up by the 
Kinetic Braiding term, $\phi'  =  \frac{1}{L_{KB}} \sqrt{\frac{r_s}{r}}$. This persists for a range of distances until 
the Horndeski term contributions in (\ref{scalareom})  catch up. That happens at the scale $r_* \sim (\frac{L_H}{L_{KB}})^{4/3} r_{3KB}$, so that for $r < r_*$ the scalar profile is again ultra-Vainshtein, $\phi' = r/L_H^2$.

\begin{figure*}[thb]
\centering
\includegraphics[height=12.5cm]{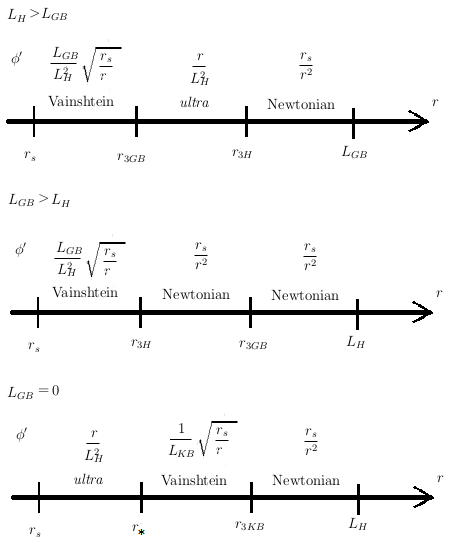}
\caption{Behavior of the H+GB theory for different values of the ratio of $L_k$s.  Here $r_\star=(L_H/L_{KB})^{4/3}r_{3KB}$.}
\end{figure*}
Our results are given in 
Figure 1. The key conclusion is that as long as the $r_3$ are greater than about $100$ AU, the `Fab Four' with the suppressed Gauss-Bonnet term has 
a phenomenologically acceptable linearized regime. This translates into the requirement that the $L_k$ are bounded as before, by 
$L_k^2 \ga 10^6 {\rm AU}^3/ {\rm km}$, or $L_k \ga 10^7$ - $10^8$ AU, although we have noticed that the Gauss-Bonnet term must be constrained more. Let us now consider more precisely just how suppressed the various terms must be.

\indent {\it Whence Schwarzschild?}: As we noted above, the Schwarzschild background is only consistent if the backreaction of the mixing terms in eq. (\ref{scalaract}) is consistently small all the way down to the Schwarzschild radius. In effect this is the same logic as adopted in studying galileons with the help of Dyson sphere sources \cite{hair}. Let us now check the constraints that the requirement for having small backreaction imposes. For starters, consider the backreaction due to the Kinetic Braiding coupling only. Explicit substitution of the Schwarzschild ansatz and scalar field profile $\phi'$, far or near, into the field equation (\ref{scalareom}) produces the error of - at most - the order ${\cal O}(r_s/r)$, showing that the solution remains consistent everywhere down to the Schwarzschild radius. In fact, exact spherically symmetric solutions found in \cite{rina}\footnote{Our Lagrangian contains terms $\sim \phi T, \partial\phi^4$, etc. that do not appear in their analysis.  These do not contribute significantly to the metric equations of motion, but do alter the scalar profile.} reduce to ours in the limit $r\ll L$, supporting our approximation schemes.  Similar results hold for pure Gauss-Bonnet\footnote{Notwithstanding the fact that Gauss-Bonnet proscribes the Vainshtein shielding.} and pure Horndeski terms, when other terms are negligible.

When all three terms are present, things can go awry, as can be seen from the reduced action with the Newtonian potential $V_N(r)$ reinstated:
\ba
S&=& 12 M_{Pl}^2 \int
drr^2\bigg\{- \frac{1}{12r^2} V_N(r^2 V_N')' -\frac12 \phi'^2 + \frac{V_N + \zeta \phi}{12 M^2_{Pl}} T  \nonumber \\
&& \, ~~~~~~~~~~~~~~~~~~~~ +L^2_{GB}\phi\Big(\frac{1}{r^2}(V_N'V_N)'+\frac{4}{r}
\phi'^3+2\phi'^4\Big) +L^2_{KB}\Big(\frac{4}{r}\phi'^3+3\phi'^4\Big) \nonumber \\
&& \, ~~~~~~~~~~~~~~~~~~~~ + L^4_{H}\Big(\frac{2}{r^2}V_N'\phi'^3+\frac{2}{r^2}\phi'^4-\frac{2}{r}
\phi'^5\Big)\bigg\} \, . ~~
\label{scalaractcorr}
\ea
Close to the source, as we have discussed before, the Gauss-Bonnet source term dominates\footnote{Unless, of course, the shortest $r_{3k}$ is shorter than the source's Schwarzschild radius.} in
determining $\phi$. This scalar field backreacts on the geometry because the mixing due to the Horndeski term scales as $L_H^4$, and is potentially large when $L_H$ is large. To estimate the backreaction we look at the distortion of the Newton's potential due to these terms, which we obtain from (\ref{scalaractcorr}). The `Fab Four'-modified equation for the Newton's potential is
\be
\frac{1}{r^2}(r^2 V_N')' =  - {\cal O}(1) \frac{L_H^4}{r^2} V_N \bigl(\phi'^3 \bigr)' +   
{\cal O}(1) \frac{L_{GB}^2}{r^2} \phi'' V_N  + \ldots \, .
\label{newteq}
\ee
Writing $V_N = - r_s/r + \delta V$, and treating (\ref{newteq}) as a perturbation equation in the expansion for $V_N$, upon 
substituting  $\phi' = ({L_{GB} r_s^{1/2}})/({L_H^2 r^{1/2}})$ we obtain
\be
\frac{(r^2 \delta V')'}{r^2} =  {\cal O}(1) \frac{ L_{GB}^3}{ L_H^2} 
\frac{r_s^{3/2}}{r^{9/2}} + \ldots \, ,
\label{newteq2}
\ee
where both leading terms on the rhs of (\ref{newteq}) contribute to the rhs of eq. (\ref{newteq2}). Integrating this gives the correction to the Newton's potential, which is $\delta V \simeq {\cal O}(1) \frac{ L_{GB}^3}{ L_H^2} \frac{r_s^{3/2}}{r^{5/2}}$. Therefore, $\delta V/V_N$ is given by
\be
\frac{\delta V}{V_N} \simeq {\cal O}(1) \frac{ L_{GB}^3}{ L_H^2} \frac{r_s^{1/2}}{r^{3/2}} \, .
\label{corrnewt}
\ee
To avoid significant corrections to the Newton's potential in the regimes where GR has been tested, the extra terms should be small.If we simply require $\delta V/V_N < 1$ to make sure that the gravitational field distortions due to the `Fab Four' extras are never excessive, and apply this to the fields of neutron stars, where $|V_N| \simeq r_s/r \simeq 1$, recalling that in this case $r_s \sim $ km, we find immediately that $L_{GB}^3 \la L_H^2 \times {\rm km}$. 

One of the most sensitive probes of GR is the lunar laser ranging. As we already mentioned the corrections of the Newton's potential which 
are neither $1/r$ nor $r^2$ lead to the precession of the orbit's perihelion. In the case of the Moon orbiting around the Earth, the current limits 
are $\delta \theta <  2 \times 10^{-11}$ per orbit \cite{lunar}. On the other hand, our formula for $\delta V/V_N$ implies
$\delta \theta \simeq \pi r \bigl[r^2[ \delta V/(rV_N)]'\bigr]' \simeq \frac{15\pi}{4} \frac{L_{GB}^3}{L_H^2} \frac{r_s^{1/2}}{r^{3/2}}$ \cite{moon}. Combining the numbers for the Earth-Moon system, this yields  
\be
L_{GB}^3 \la L_H^2 \times 0.1 {\rm km} \, .
\label{gbbound}
\ee
The numerics are only slightly better than the neutron star bound above. Nevertheless these actually are fairly strong bounds especially when combine them with the strong coupling limits of the next section. At any rate, already at this point we can conclude that the Gauss-Bonnet term can't play any significant role at cosmological scales (nonlinear effects notwithstanding, to be discussed later on). 

Note that in the above discussion the Newtonian potential never receives direct corrections from backreaction fom the Kinetic Braiding term in the leading order. These results seem to hold even if
$L_{KB} \gg L_H$, however we should still check the interplay of the Kinetic Braiding and Horndeski terms at short distances. As it turns out these are not a problem. Indeed, let $L_{KB} \gg L_H$, and let Gauss-Bonnet be negligible all the way down to the gravitational radius of the mass source. In this case, the scalar field profile near the source, for $L_H$ bounded by above discussion, is $\phi' = r/L_H^2$; as long as
$L_H \ga 10^7$ AU, the direct force which the scalar induces at large distances is within bounds, as we showed previously. The backreaction of this scalar profile on the Newton's potential turns out to be automatically small:
eq. (\ref{newteq}) is still the same, but upon substituting the scalar profile $\phi' = r/L_H^2$ and 
$V_N = - r_s/r + \delta V$, we obtain
\be
\frac{(r^2 \delta V')'}{r^2} =  {\cal O}(1) \frac{ r_s}{ L_H^2 r} + \ldots \, ,
\label{newteq3}
\ee
This yields a correction to Newton's potential $\delta V \simeq {\cal O}(1) \frac{r_s r}{ L_H^2}$, and so $\delta V/V_N$ is 
\be
\frac{\delta V}{V_N} \simeq {\cal O}(1) \frac{r^2}{ L_H^2} \, .
\label{corrnewt2}
\ee
Using lunar laser ranging to constrain this, we now find $\delta \theta \simeq \pi r \bigl[r^2[ V/(rV_N)]'\bigr]' \simeq 2\pi \frac{r^2}{L_H^2}$. Using $\delta \theta < 2 \times 10^{-11}$ per orbit for the Earth-Moon system yields 
$L_H \ga  10^{12} \, {\rm km} \sim 10^4$ AU. In fact, this is not surprising - the scalar correction to the long range potential, $\sim r^2/L_H^2$ is clearly greater than $r_s r/L_H^2$, and so once it is under control the backreaction is automatically small.

So, in sum we see that the classical bounds from Solar system physics strongly constrain the Gauss-Bonnet term, require the Horndeski term coefficient to be $L_H \ga 10^7$ AU, but appear to favor a large Kinetic Braiding coefficient.

\section{Perturbations and the Strong Coupling Cutoff}

So far we have been exploring the bounds on `Fab Fours' from the requirement that the classical theory is consistent with observational tests of
gravity, and specifically the Newtonian limit. Let us now explore the bounds from the validity of quantum perturbation theory to the leading order, and specifically the issue of the strong coupling in perturbation theory. Because the theory has derivative couplings, the scattering amplitude 
of the scalar modes saturates the unitarity bounds at a scale much below $M_{Pl}$. To find where it happens, one can follow the by-now standard approach from galileon models, compute the 
tree level $2\rightarrow 2$ scattering amplitude of $\phi$ modes, and look for the scales where the amplitude becomes of order unity.
The idea is to write down the perturbative scalar Lagrangian as 
${\cal L}_{scalar} = - Z^\mu{}_\nu \delta \phi_\mu \delta \phi^\nu + {\rm interactions}$, canonically normalize the fluctuations $\delta \phi$, and then using the couplings for the canonically normalized fields, compute the scattering amplitudes. 

Perturbative expansion of the theory (\ref{canact}) 
about the classical background set by a mass source of a given $r_s$ yields for the quadratic kinetic matrix of the theory the following expression:
\ba
Z^\mu_\nu &=& \delta^\mu_\nu+8L_{KB}^2\texttt{diag}\bigg(\phi''+\frac{2}{r}\phi',\frac{-2}{r}\phi',\phi''+\frac{1}{r}\phi',\phi''+\frac{1}{r
}\phi'\bigg)\nonumber\\
&+& L_{GB}^2\texttt{diag}\bigg(6\phi''+\frac{12}{r}\phi',9\phi''+\frac{12}{r}\phi',6\phi''+\frac{15}{r}\phi',
6\phi''+\frac{15}{r}\phi'\bigg)\nonumber\\
&+& 3L_{H}^4\texttt{diag}\bigg(\frac{4}{r}\phi'\phi''+\frac{2}{r^2}\phi'^2,\frac{2}{r^2}\phi'^2,\frac{2}{r}\phi'\phi'',\frac{2}{r}\phi'\phi
''\bigg)\Bigg(1+\mathcal{O}\bigg(\frac{r_s}{r},\frac{r^2}{L_H^2}\bigg)\Bigg) \, .
\label{Zs}
\ea
This matrix, while diagonal, is clearly anisotropic. Nevertheless, the eigenvalues are all comparable in the regimes we are interested in, 
very close to the sources. The solutions do not support the types of pathological anisotropies encountered in $4D$ massive gravity \cite{drgt,bkp}. 
So in what follows we will simply ignore the anisotropies in (\ref{Zs}) and model the $Z$-matrix as a constant matrix, with scale
$Z = 1 + Z_{GB} + Z_{KB} + Z_{H}$. From (\ref{Zs}), we see that
\be
Z_{GB} = {\cal O}(1) L_{GB}^2 \frac{\phi'}{r} \, , ~~~~~~~~~~~ Z_{KB} = {\cal O}(1) L_{KB}^2 \frac{\phi'}{r} \, , ~~~~~~~~~~~ Z_{H} = {\cal O}(1) L_{H}^4 \bigg(\frac{\phi'}{r}\bigg)^2 \,. 
\label{eigens}
\ee
The specific expressions for these Zs can be evaluated in the various regimes of interest.

\begin{figure*}
\centering
\includegraphics[height=4cm]{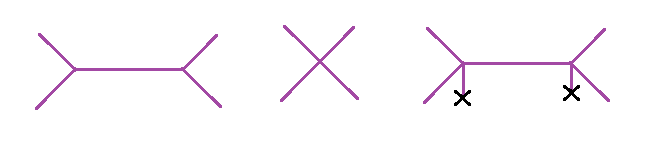}
\caption{The three types of contributions to the $2\rightarrow 2$ scattering. The first diagram is dominant for the Kinetic
Braiding and Gauss-Bonnet insertions, while one of the other two dominates for the Horndeski term, depending on the
region of parameter space.}
\label{4point}
\end{figure*}

Let us next demonstrate the technique by considering the interactions mediated by the Kinetic Braiding term. In this case the expansion 
of (\ref{canact}), using $\phi=\bar{\phi}+\sigma/\sqrt{Z_{KB}}M_{Pl}$ and the Schwarzschild metric, with $Z_{KB} = L_{KB} r_s^{1/2}/r^{3/2}$, yields
\be
\mathcal{L}_{KB}=-\frac{1}{2}\partial\sigma^2+\frac{L_{KB}^2}{\sqrt{Z_{KB}}^3M_p}\big(2\partial\sigma^2\Box\sigma-2\sigma_{\mu\nu}\sigma^\mu
\sigma^\nu+12\bar{\phi}_\mu\sigma^\mu\partial\sigma^2\big)+\frac{3L_{KB}^2}{Z_{KB}^2M_p^2}\partial\sigma^4 \, .
\label{pertsL}
\ee
Based on this (see Fig. (\ref{4point})), we can construct four different processes. The first tree level scattering process $2 \rightarrow 2$ involves 
two cubic vertices and an internal propagator, so that the vertices each contribute $p^4$, and the propagator $p^{-2}$ to the 
momentum transfer. The diagram evaluates to $\mathcal
A\sim(L_{KB}^2/Z_{KB}^{3/2}M_p)^2p^8/p^2=\ell^6p^6$, with $\ell=L_{KB}^{2/3}l_p^{1/3}/Z_{KB}^{1/2}$. The diagram with both vertices
attached to the background field has an additional contribution to the prefactor proportional to $\phi'^2$ and two fewer powers of
momenta, yielding $\mathcal
A \sim \hat \ell^4p^4$ with $\hat \ell=\sqrt{\frac{r_{KB3}\phi'}{Z_{KB}^{1/2} \ell}} < \ell$ even for the
most dangerous estimation of $\phi'$. The process involving only one contraction with the external field yields a geometric mean of the two amplitudes, and so its cutoff scale is in between these two cases. Finally, the 4-point vertex also has quartic momentum dependence, yielding $\mathcal
A \sim \bar \ell^4 p^4$, with $\bar \ell =\sqrt{L_{KB}l_p/Z}=(l_p/L_{KB})^{1/6} \ell \ll \ell$. So the first process dominates in this class,
giving for the UV cutoff of the theory the scale
\be
\ell= \frac{L_{KB}^{2/3}l_p^{1/3}}{Z_{KB}^{1/2}} = \frac{L_{KB}^{1/6} l_p^{1/3} r^{3/4}}{r_s^{1/4}} \, .
\label{KBcutoff}
\ee
Applying this to terrestrial conditions, with $r_s \sim10^{-5}$ km and $r \sim 6000$ km, 
we find $\ell = 10^{-3} (L_{KB}/{\rm km})^{1/6}$ mm, which for $L_{KB} \sim H_0^{-1} \sim 10^{23}$
km is $\ell \simeq$ few cm, just like in galileon models \cite{gal}. This may seem problematic on one hand. However it is relatively close to the scales of tabletop experiments, 
and perhaps the effects which lie beyond the cutoff do not build up sufficiently quickly to run afoul of the experimental limits \cite{gal}. If we insist on pushing the strong coupling scale down to about $1$ mm, however, the Kinetic Braiding coupling scale should be about 9-10 orders of magnitude smaller, $L_{KB} \la 10^6$ AU. 

The Gauss-Bonnet sector of fluctuations is very similar to the Kinetic Braiding sector, and so similar results follow. The crossover processes which involve the vertices from different sectors yield the cutoffs which are geometric
means of the two length scales, and so are always negligible. However, since the Vainshtein shielding never sets in, the background fields are not very good in protecting the scalar fluctuations from unitarity loss at low scales. The Horndeski Lagrangian is more intricate. The dominant channel changes with $Z_{H}$, and therefore $r$, and also with the
ratios of $L_k$s. The most important processes yield $\mathcal{A}\sim L_H^{8/3}l_p^{4/3}p^4/Z^{4/3}$ (the third diagram of
Fig. \ref{4point}), and $\mathcal{A}_2\sim L_H^{16/3}r_s^{4/3}l_p^{4/3}p^4/Z^{4/3}r^4$ (the second diagram), which yield $\ell_H = \frac{L_{H}^{2/3}l_p^{1/3}}{Z^{1/3}}$ or  $\ell_H = \frac{L_{H}^{4/3}r_s^{1/3}l_p^{1/3}}{r\sqrt{Z}}$ for the strong coupling scales in the Horndeski sector, depending on the background.

\begin{figure*}[th]
\centering
\includegraphics[height=8cm]{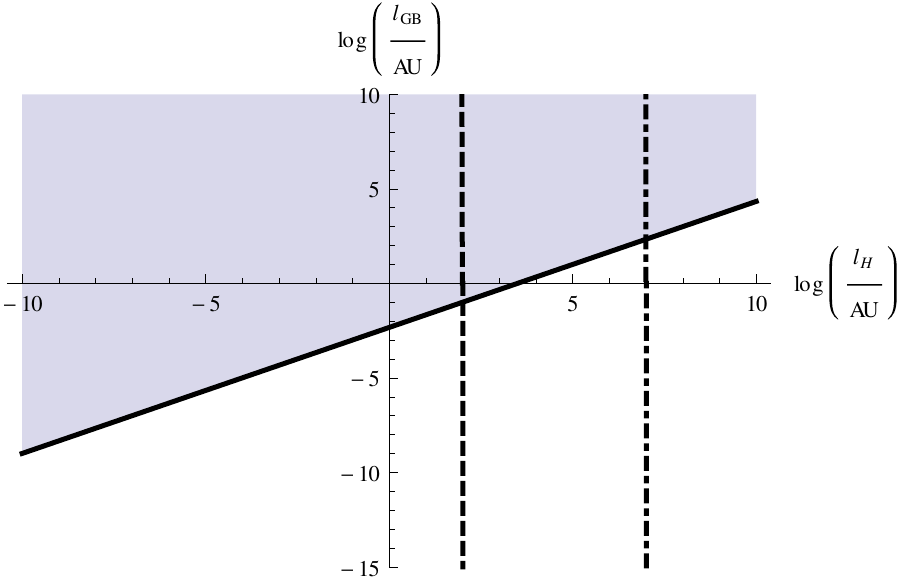}
\caption{Constraints on the Gauss-Bonnet and Horndeski length scales. ÊThe region above and to the left of the dashed line is excluded. The solid line comes from lunar laser ranging bounds. ÊThe dot-dashed line has a strong coupling scale of $\sim 10$ cm, whereas the dashed line has a strong coupling
scale of $\sim 1$ mm.}
\label{ruledout}
\end{figure*}

The main conclusion is that the scales which control the scalar exchange processes are 
\begin{equation}
\ell_{UV} = \Bigg(\frac{L_{KB}^{2/3}l_p^{1/3}}{\sqrt{Z}},\frac{L_{GB}^{2/3}l_p^{1/3}}{\sqrt{Z}},\frac{L_{H}^{2/3}l_p^{1/3}}{Z^{1/3}},
\frac{L_{H}^{4/3}r_s^{1/3}l_p^{1/3}}{r\sqrt{Z}}\Bigg) \, ,
\label{scalesz}
\end{equation}
where $Z$ is set by the dominant short distance classical solution for $\phi$. It is 
the largest of the eigenvalues of (\ref{Zs}). Bounds can
therefore be placed on the largest of the length scales in (\ref{scalesz}), completely analogously to the Kinetic Braiding example above. The specific numerical expressions depend on the background and the combination 
of terms involved, as we will illustrate with several examples. Several of these are depicted in the diagram Fig. \ref{ruledout}.

Let us consider the case where the Kinetic Braiding is not the dominant coupling. Then, we 
keep the Gauss-Bonnet term, but impose $L_{GB} \le 1$ AU to satisfy the classical constraints. The near-field solution $\phi' = \frac{L_{GB}}{L_H^2} \sqrt{r_s/r}$ in terrestrial conditions yields $Z = \frac{L_{GB}^2 r_s}{r^3} \simeq 1$, 
independently of $L_H$. This solution therefore does not help with improving the short distance cutoff, which is fully controlled by $L_H$.
If we now take $L_H \sim H_0^{-1} \sim 10^{23}$ km, the cutoff would be as low as $\ell_{UV} \simeq 100$ km, which is very low. If we instead take 
$L_H \ga 10^7$ AU, to ensure that the Solar system bounds are met, the cutoff becomes $\ell_{UV} \simeq {\rm few} \times 10$ cm. Insisting that it is as low as a millimeter requires the Horndeski length scale $L_H$ to be lowered down to about $100$ AU.

\begin{figure*}[thb]
\centering
\includegraphics[height=10cm]{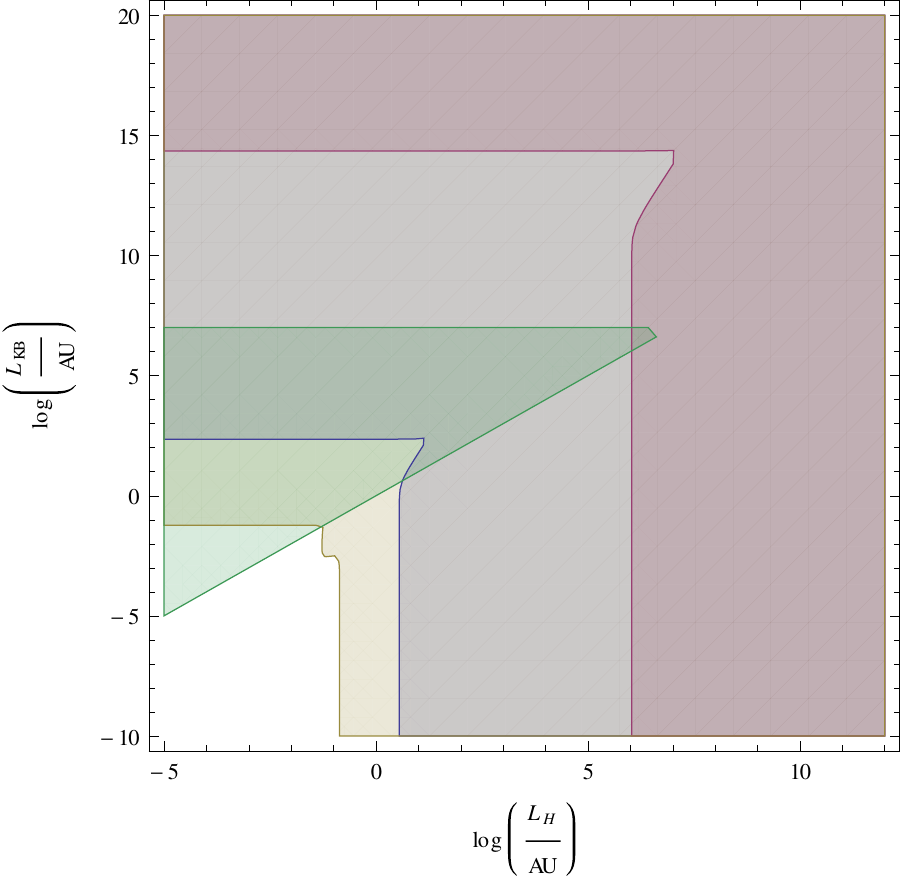}
\caption{
The blue region excludes regions of the parameter space where $\ell > {\rm mm}$, while the pink region relaxes the bound to $10 {\rm cm}$ and the yellow strengthens it to $0.1{\rm mm}$. ÊThe green
region excludes places where the Vainshtein radius $r_v<100 {\rm AU}$ for comparison. Ê}
\label{allowed}
\end{figure*}

Finally let us consider the interplay of the Kinetic Braiding and Horndeski terms. If $L_{H}>L_{KB}$, the background features the 
ultra-Vainshtein effect. Again the perturbations are {\it not} shielded: $Z_H \sim L_H^2 \phi'^2/r^2 \simeq 1$.
So the quantum fluctuations saturate the unitarity bound at the scale
$\ell_{UV} \simeq L_H^{2/3} l_p^{1/3}$, just like in the Horndeski-Gauss-Bonnet case above: for 
$L_H \ga 10^7$ AU, the cutoff length is again about $\ell \simeq {\rm few} \times 10$ cm. And again, 
to push the cutoff length 
down to mm, the Horndeski length scale should be $L_H \le 100$ AU.  If, on the other hand $L_{H} < L_{KB}$, the bounds depend on 
whether the Earth's orbit sits in the regime where the ultra-Vainshtein profile has kicked in or not. The crossover happens at the
scale $r_\star=(L_{H}/L_{KB})^{4/3}r_{3KB}$. Whether we are above or below this transition is determined by whether
$L_{H}^2/L_{KB}\lessgtr10^{-4}AU$. If we are not orbiting at an ultra-Vainshtein radius then
we get the additional constraint that $L_{KB}\lesssim 100  AU$, which for our liberal interpretation of current constraints is
not as strong a bound as demanding that the Vainshtein radius lie completely outside the solar system, $r_V<100 AU$. However, if $L_{KB} \gg L_H$, the near field profile features the ultra-Vainshtein form, 
and so the $Z$ factor in (\ref{scalesz}) is dominated by the Kinetic Braiding contribution from (\ref{eigens}),
$Z \simeq L^2_{KB}/L_H^2 \gg 1$. This is in fact the best case scenario for the `Fab Four': the strong coupling scale is
\be
\ell_{UV} \simeq L_H \Bigl(\frac{l_p}{L_{KB}} \Bigr)^{1/3} \, ,
\label{bestuv}
\ee
which for $L_H \ga 10^7$ AU and $L_{KB} \sim H_0^{-1}$, which satisfy all classical constraints, numerically is
$\ell_{UV} \simeq 1$ cm, again just like the galileons \cite{gal}. 

So the `Fab Four' models where
$L_{GB} \ll L_H \ll L_{KB}$ might -- very marginally -- pass the Solar system bounds, potentially producing something interesting in the table top experiment range, and yet have some effects at cosmological distances. A summary of these results is given in Fig. \ref{allowed}.

\section{Summary}

`Fab Four' models are a subset of nontrivial scalar-tensor extensions of Brans-Dicke models which involve derivative couplings but are unitary, in the sense that all the propagating modes obey the standard Cauchy problem and are not ghosts. In this work we have explored phenomenological constraints on them. Requiring that the Newton's law is correctly reproduced at large distances (ie within the Solar system where gravity has been tested extensively), and focusing on a broad class of models where the coupling functions are approximated by the leading order constants or linear terms we have looked for the conditions where the Vainshtein shielding can tame the extra scalar force. Further, we have checked the validity of the quantum theory of scalar fluctuations on top of the background, and determined the scales at which the theory becomes strongly coupled, and ceases to be predictive. 

Our conclusions are that the Gauss-Bonnet term must be significantly suppressed relative to the other `extras' in the `Fab Four' framework, because it sources the largest distortions of the Newton's potential and obstructs the Vainshtein shield from setting in. Since we have been working in the limit where we truncated the `Fab Four' coupling functions to the leading nontrivial order, this really means that the Gauss-Bonnet coupling function must have a very small coefficient of the linear term in its Taylor expansion, in order to avoid violating observational bounds.  Other `extras' behave better, allowing the length scales that control them to be considerably larger, $L_H \ga 10^7$ AU, and $L_{KB} \sim H_0^{-1}$, passing the Solar system constraint and marginally meeting the unitarity bounds, while allowing for the possibility of some horizon scale effects. The main problem is that the theory generically loses predictivity at length scales of the order of few cm to few tens cm, like the galileon models \cite{gal}. One might hope that the relatively small gap between the cutoff and the currently probed regimes of gravity might not be enough to dramatically violate the observation. On the other hand if one demands perturbativity down to a millimeter, the length scales are generically bound to be small, so the `Fab Four' effects would be confined to the Solar system. We stress that the bounds could be relaxed by either adding extra terms $\propto (\partial \phi)^2A$, skipping leading order constant and linear terms in the coupling function expansions (in which case, the leading order effects will generically be too small) and/or involving backgrounds with very large scalar derivatives (which could yield much stronger backreaction on he Newton's potential). In any case, the framework may provide useful straw men for future tests of General Relativity, in the lab and on the sky. Further explorations, specifically of relativistic corrections to the Newtonian limit, may therefore be warranted.

\vskip.5cm

{\bf \noindent Acknowledgements}

\smallskip

We would like to thank Tony ``The Padilla''  Scouser and Norihiro Tanahashi for useful discussions.
NK thanks the School of Physics and Astronomy, U. of Nottingham for hospitality in the course of this work.
This work is supported by the DOE Grant DE-FG03-91ER40674. NK is also supported 
by a Leverhulme visiting professorship.

\end{document}